\documentstyle[11pt,aaspp4]{article}

\slugcomment{To appear in Astrophysical Journal (Letters).}

\lefthead{Yun et al.}
\righthead{Resolution and Kinematics of Molecular Gas 
Surrounding the Cloverleaf Quasar at z=2.6 
Using the Magnifying Property of the Gravitational Lensing}

\newbox\grsign \setbox\grsign=\hbox{$>$} \newdimen\grdimen 
\grdimen=\ht\grsign
\newbox\laxbox \newbox\gaxbox
\setbox\gaxbox=\hbox{\raise.5ex\hbox{$>$}\llap
     {\lower.5ex\hbox{$\sim$}}}\ht1=\grdimen\dp1=0pt
\setbox\laxbox=\hbox{\raise.5ex\hbox{$<$}\llap
     {\lower.5ex\hbox{$\sim$}}}\ht2=\grdimen\dp2=0pt

\def\kms    {\ifmmode{{\rm ~km~s}^{-1}}\else{~km~s$^{-1}$}\fi}

\def\h2     {H$_2$}

\newbox\grsign \setbox\grsign=\hbox{$>$} \newdimen\grdimen 
\grdimen=\ht\grsign
\newbox\simlessbox \newbox\simgreatbox
\setbox\simgreatbox=\hbox{\raise.5ex\hbox{$>$}\llap
     {\lower.5ex\hbox{$\sim$}}}\ht1=\grdimen\dp1=0pt
\setbox\simlessbox=\hbox{\raise.5ex\hbox{$<$}\llap
     {\lower.5ex\hbox{$\sim$}}}\ht2=\grdimen\dp2=0pt

\def\simless{\mathrel{\copy\simlessbox}}
\def\go{
\mathrel{\raise.3ex\hbox{$>$}\mkern-14mu\lower0.6ex\hbox{$\sim$}}
}
\def\lo{
\mathrel{\raise.3ex\hbox{$<$}\mkern-14mu\lower0.6ex\hbox{$\sim$}}
}

\begin{document}

\title{Resolution and Kinematics of Molecular Gas Surrounding the
Cloverleaf Quasar at z=2.6 Using the Gravitational Lens}

\author{M. S. Yun\altaffilmark{1}, N. Z. Scoville, J. J. Carrasco,
\& R. D. Blandford}
\affil{California Institute of Technology, 
    Pasadena, CA 91125}

\altaffiltext{1}{present address: National Radio Astronomy Observatory,
	P.O. Box 0, Soccoro, NM 87801, USA ({\it myun@nrao.edu})} 
 
\begin{abstract}

Gravitational lenses have long been advertised as primitive 
telescopes, capable of magnifying cosmologically distant sources
(\cite{zwi37}).
In this Letter, we present new, $0''.9$ resolution CO~(7--6) observations 
of the $z=2.56$ Cloverleaf quasar (H 1413+117) and spatially resolved
images.  By modeling the gravitational lens, we infer a size scale of $0''.3$ 
($\sim1$~kpc) for the molecular gas structure surrounding the quasar, and
the gas has a kinematic structure roughly consistent with a rotating disk.
The observed properties of the CO emitting gas are similar to
the nuclear starburst complexes found in the infrared luminous galaxies
in the local universe, and metal enrichment by vigorous
star formation within this massive nuclear gas complex can explain
the abundance of carbon and oxygen in the interstellar medium of this
system observed when the universe was only a few billion years old.
Obtaining corresponding details in an unlensed 
object at similar distances would be well beyond the reach of 
current instruments, and this study highlights the less exploited
yet powerful use of a gravitational lens as a natural telescope. 

\end{abstract}


\section{Introduction}

The Cloverleaf quasar (H 1413+117)
is one the most distant objects where the presence of a 
large amount of cold, molecular gas is inferred from the detection of bright 
CO emission (\cite{bar94}).
Its spectral energy distribution peaks in the far-infrared as in 
another luminous high redshift object FSC~10214+4724 (z=2.29)
(\cite{bar92,row93,eis96}), and the total infrared luminosity and inferred 
molecular gas mass are comparable for these two objects.  The optical 
image consists of four clearly separated peaks implying that the (unseen)
lens lies close to the line of sight to the quasar (\cite{kay90}).
While the optical quasar may appear unresolved, the surrounding gas
and dust is expected to be substantially extended, and the new
CO~(7--6) observations are made at $1''$ resolution in order to 
resolve the distribution of molecular gas complex surrounding 
the quasar.  Previous CO observations at much lower spatial
resolutions ($\theta=$ 3$''$-8$''$) had failed to resolve the molecular
gas structure (\cite{bar94,wil95}).

In the following sections, we show that the CO~(7--6) emission 
in the Cloverleaf quasar is indeed resolved by our new interferometric
observations and demonstrate that 
$0''.1$ scale spatial details can be revealed by gravitational lensing
in addition to the usual luminosity amplification.  This magnifying
property of gravitational lens holds promise for probing
minute details of distant objects unachievable through a
conventional observational approaches.

\section{Observations and results}

Aperture synthesis observations of CO (7--6) emission 
($\nu_0=806.651776$~GHz)
redshifted to 226.70~GHz were carried out at the Owens Valley Millimeter
array between December 1994 and 
March 1995 in the equatorial and high resolution 
configurations with longest baselines 200~m (E-W) and 220~m (N-S).
Typical system temperatures 
were 500-800~K (SSB), mainly contributed by the atmosphere. A digital 
correlator configured with $120\times4$~MHz (5.3~km s$^{-1}$) channels
covered a total velocity range of 636~km s$^{-1}$. 
The data were calibrated using the standard Owens Valley
array program MMA (\cite{sco92}), and the continuum and
spectral line maps are made with DIFMAP (\cite{she94}).  
Data from total of 30 baseline pairs cover the 
inner 12-180 k$\lambda$ uv-plane, and the resulting synthesized 
beam is $0''.86\times1''.0$ (PA$=73^\circ$)
with natural weighting of the visibility data.  The uv-coverage is
quite uniform, and synthesized beam is well-behaved -- the largest
sidelobes appear 2$''$ north and south of the main beam with 
20\% of the peak amplitude.  The CLEANing of the maps was limited 
only to the central 3$''$ diameter region - no emission is seen elsewhere
in the ``dirty" maps.  The positional accuracy of
the resulting maps is better than $0''.2$. After 25 hours of on-source
integration, we achieved 3.2 (LSB) and 4.9 (USB) mJy beam$^{-1}$ sensitivity 
in the 1~GHz bandwidth analog correlator continuum data. 
The CO (7--6) emission appears in only 1/2 of the total 120
spectrometer channels,
and 1$\sigma$ noise in each 110~MHz ($\Delta V=145$~km s$^{-1}$)
averaged channel maps (Figures 1b \& 1c) is 8.3~mJy beam$^{-1}$. 
No continuum emission is detected in the 230 GHz USB continuum data,
and the $3\sigma$ upper limit of 15 mJy beam$^{-1}$ is marginally 
consistent with the 240 GHz continuum flux of $18\pm2$ mJy 
reported by Barvainis et al. (1995).  The summary
of the observations and results are given in Table 1.

The new CO~(7--6) map resolves the source for the first time (Fig. 1).  
The velocity integrated CO emission, Figure 1a, approximately 
follows the optical light suggesting that
the CO source is close to the quasar.  However, the peak of the CO emission 
is located near image B, whereas image A is the brightest in the 
optical and infrared -- perhaps a first hint that the distribution 
of the CO emission 
is different from that of the quasar light.  Since the CO emission does 
not form an Einstein ring (c.f., MG 1131+0456, \cite{hew88}),  
the CO emitting region must be smaller in size than that
of the tangential caustic, $\sim 1''$ or about 4 $h^{-1}$ kpc 
(assuming that $\Omega_\circ =1$).  Our integrated CO (7--6)
spectrum has the same profile as the IRAM 30-m spectrum (R.
Barvainis, private communication), and it is not shown because
of the space limitation in this {\it Letter}.

In order to study the kinematics of the gas, we produced separate images
of the blue- and redshifted gas (Fig. 1c \& 1d).  Clear 
differences from the optical image are seen in both the blue-
and the redshifted images while the difference between the two is
also significant as shown in Fig. 1b.  Thus two 
kinematically distinct sources have been spatially resolved by the 
gravitational lens.  As in the velocity integrated image (Fig. 1a),
the blue- and redshifted gas images also have multiple bright
peaks but spatially displaced from each other and from the optical quasar 
positions.   This displacement is quite significant,
each $\sim$7 times the rms noise in the maps.  Because the
position information depends on the phase of the visibilities, 
which is measured much more accurately than amplitude by an 
interferometer, the significance of the offset is even greater. 
In addition, the fact that the multiple peaks in both blue and redshifted 
CO images can be mapped back to single distinct (but different) features
in the source plane further supports the reality of observed
differences (see below).

Comparison of our data with unpublished Plateau de Bure 
(PdB) interferometer
maps (R. Antonucci \& R. Barvainis, private communication) 
also lends support to the extended nature of the CO emission.  
The higher resolution PdB maps show a 3-image geometry for the CO
emission, similar to our own, but observed differences as
function of velocity are
much less pronounced.  The total line flux mapped by the PdB 
interferometer is about 25\% less than our measurement, and their
uv-coverage poorly samples the baselines shorter than 50-m in length 
(needed to map 2-3$''$ scale structures).  
Only 30\% of the total flux is seen in the difference map (Fig. 1b), 
and thus one may infer that about 70\% of the line flux orignates 
in a very compact region, very close to the quasar, while the
remaining 30\% lies further out, if the both observations
are accurate.

\section{Gravitational Lensing Model and Inferences}

In order to test our hypothesis that the observed difference between
the blue and red CO images is due to spatial resolution of the gas
distribution and kinematics, we have developed a gravitational 
lens model.   We have modeled the lens as a single
elliptical potential with an external shear (\cite{sch92})
$$
\psi=b[1+(1-\gamma_c)(x/s)^2-2\gamma_s (xy/s^2)+(1+\gamma_c)(y/s)^2]^q-
\gamma_1(x^2-y^2)-2\gamma_2xy
$$
where $b$ measures the size of the Einstein ring and $s$ is a core
radius.   The parameters $\gamma_{c,s}$ represent the 
ellipticity in the galaxy
potential and the parameters $\gamma_{1,2}$ measure the strength of the
tidal action of external galaxies.  The exponent $q$ is dictated by the
radial density profile.  A value $q=0.5$ corresponds to the
(pseudo-)isothermal case.   We solved for parameters which reproduce 
the HST image locations (\cite{fal93})
and which also recover the three relative magnifications of the four 
K band images measured nearly contemporaneously
with the Keck telescope (M. Pahre, private communication).  
These relative magnifications are quite similar to their optical
counterparts.  We take this as an indication that microlensing is
unimportant. For an emitted wavelength $\sim600$~nm, a thermal source
would be too large to be treated as a point source when microlensed by
$\sim0.1$~M$_\odot$ stars in an intervening galaxy (cf. \cite{rau91}). A non-thermal, red source could, however, be subject
to strong, achromatic microlensing and if this is happening, 
our model would definitely be
invalid. High accuracy photometric monitoring may determine if
microlensing is at work. 
(In a different approach to modeling this source, Kayser et al (1990)
invoke microlensing and ignore the measured magnifications in deriving a
macrolensing model.  They present two quite different models that fit the
image locations; one involves a single 
galaxy with eccentricity larger than 0.9, the other uses two
galaxies.  Neither model is compatible with our CO data -- see below.)

Our model is able to reproduce both the image locations and relative
fluxes to within the measurement accuracies.  The best fitting parameters
are $b=0.088''^{{}^2}, ~s=0.21'',~\gamma_c=0.09, ~\gamma_s=0.15, ~q=0.61, ~\gamma_1=-0.05, ~\gamma_2=-0.08$.  The lensing galaxy is centered at 
($-0.13''$, +$0.56''$) relative to image A.  The 
corresponding ellipticity in the surface density contours is $\epsilon=0.3$,
and the position angle of its major axis is $125^\circ$ 
(see Figure 2a).   This model is quite different from the single
galaxy model of Kayser et al.  It is roughly orthogonal in orientation and
much less elliptical. The total point source magnification $\mu=220$ is
much larger than the Kayser et al model implying that the source is
correspondingly less powerful. It also implies that the time delays are
very much shorter.  Adopting $z_d=\Omega_0=1$, for example, we obtain
$t_{BA}=-0.3,~t_{CA}=-0.5,~t_{DA}=0.9~h^{-1}$~day, which will 
be a challenge to measure.  We have preferred
this model because it seems best able to accommodate the CO data.
However, as the foregoing discussion implies, it is not unique and
much more will need to be known before this source becomes a good
candidate for determining the Hubble constant.

We now adopt this lens model and use it to derive CO source 
models that provide the best fit to the red and the blue images.  Using the adopted lens model, we were able to produce
an excellent fit to the contours of the redshifted CO image; the fit to the 
blueshifted image was less good but acceptable granted the resolution of the 
observation (Fig. 2b \& 2c). We find that the peaks of the red and the blue source are displaced by $\sim0''.15 \equiv 550 h^{-1}$~pc
with respect to the quasar, roughly on opposite sides of the optical source.   
This suggests that the molecular gas orbits the quasar.
We obtain a dynamical estimate of the central mass of 
$\sim10^{10}h^{-1}$~M$_\odot$ adopting a low inclination $i\sim30^\circ$, on
the grounds that H~1413+117 is a broad absorption line (BAL) quasar.
(Note, though, that the quasar is not being observed through the molecular gas
as the high extinction $A_v > 1000$ would render it optically invisible.) 
The fact that each of the multiply imaged  CO sources, each with 
$\sim 7\sigma$ in significance, map back to a {\it single} image 
on opposite side of the quasar gives us additional confidence in our 
modeling and the reality of the observed
differences between the red and blue images.
The two CO sources are not exactly symmetric about the
quasar, and this may reflect a real asymmetry in the gas 
distribution, an error in relative astrometry between optical and CO
images ($\simless 0.''2$), or an inaccuracy in the lens model. 

The overall CO magnification depends on the source size (\cite{eis96}). 
For our predicted overall CO source size, we
estimate that the mean magnification is $\mu_{CO}\sim10$. A similar
magnification is probably appropriate for the far infrared emission
discussed below.  Adopting this value, the associated molecular 
gas mass is calculated from the observed
flux ($S_{CO(7-6)}=41$ Jy km s$^{-1}$) 
to be $\sim2\times 10^{10} h^{-2}$~M$_\odot$ using a standard
Galactic CO-to-H$_2$ conversion factor (\cite{you91,sol92}).
A lower than solar metallicity requires a somewhat larger 
gas mass to account for
the observed CO flux.  Because of the small source size ($\simless 3''$), 
our interferometer measurement should include all the emission --
the existing single dish measurement of CO (7--6) flux is 
quite uncertain because it suffers from lack of a good spectral baseline
(R. Barvainis, private communication).
This inferred mass of molecular gas is comparable with
the upper bound derived on the basis of dynamics, and we therefore
suppose that it is a fair estimate of the total molecular hydrogen mass.
This in turn implies that molecular gas may dominate the mass 
distribution on the $\sim1$ kpc scale just as
in the ultraluminous infrared galaxies in the local universe (\cite{sco91b,yun95,bry96}).

\section{Physical Properties of the Molecular Gas Surrounding the Quasar}

The CO emitting gas appears to be both warm and dense.  The peak observed 
brightness temperature for the CO~(7--6) line seen in the  
individual 32~MHz  channel maps is 2.2~K above 
the brightness temperature of the cosmic background radiation. 
Gravitational lensing preserves brightness temperature, and this excess 
corresponds to a Planck brightness temperature at the source redshift
of $24$~K.  If the lensed image fills 1/2 (1/4) of the synthesized beam,
then the inferred temperature of the gas is about 50 (100) K.  
The CO (7--6) line is bright, and the near unity ($\sim 0.9$) inferred 
brightness temperature ratio of our CO (7--6) measurement
to that of the CO (3--2) measurement by Barvainis et al. (1994)
suggests that the CO emitting gas is thermalized and 
optically thick if both transitions originate from the same region.
(In support of this, we note that both transitions have the same 
line width and profile.)  Because the CO J = 7 rotational level lies 
155 K above the ground state, producing significant CO (7--6) emission 
and the observed line ratio requires a minimum  
density of $n>10^5$ cm$^{-3}$ and an excitation temperature $T_{ex}\go 100$ K.  
The inferred dust temperature from the continuum spectrum is
$\sim100$~K (\cite{bar92,dow92,row93}), and this is consistent with
the expectation that the dust and gas should be in thermal equilibrium
at the density required for producing the observed CO (7--6) flux.
The observed
spectral energy distribution for the infrared and submillimeter emission 
can be interpreted as thermal emission from dust with total mass of $\sim5\times10^7$~M$_\odot$, which 
corresponds to a gas mass $2.5\times10^{10}$~M$_\odot$ after
correcting for the lens magnification and adopting a gas
to dust ratio $\sim500$ appropriate to infrared galaxies in the local 
universe (\cite{san91}).  This is in good agreement with the gas mass
estimate from the CO emission above although large uncertainties
are associated with both estimates.  

The infrared luminosity to 
molecular gas ratio $L_{FIR}/M_{H_2}$ is a crude measure of the 
efficiency of converting gas into radiant energy either through stars
or via an active nucleus.  Among the most luminous infrared
galaxies found in the local universe, 
this ratio approaches $\sim100$~L$_\odot/$M$_\odot$.  However, for  
H~1413+117 and FSC~10214+4724, the observed ratios exceed 
200~L$_\odot/$M$_\odot$ (neglecting the possibility of different
magnification for the infrared and CO emission).  This suggests 
that the bulk of the large infrared luminosities of 
both of these objects is provided 
by a quasar, that is hidden in the case of FSC~10214+4724.  
In these discussions, it is assumed that the CO emitting 
region coincides with the infrared emitting region because 
a large amount of dust associated with the CO emitting dense
molecular gas would dominate the total infrared emission.  

In summary, we note that the molecular gas complex in H~1413+117 
is similar in size, mass, density, and temperature, to that 
observed in the nuclei of local infrared galaxies (\cite{san91}).
This is consistent with theoretical models in which a large quantity of 
gas driven into the nuclear region by mergers or interactions (\cite{bar91})
is processed by starbursts over a time scale $\sim10^8$ years
so that it builds up a high metallicity (Scoville \& Soifer 1991).
This process may account for the strong CO emission from the ISM of 
this z=2.6 system observed when the
universe was only a few billion years old. 

Here we have also demonstrated that gravitational lens indeed 
can be a valuable 
tools for probing objects at cosmological distances by yielding details
that are be normally inaccessible for the current generation of observational instruments.  For example, the resolution of the 
distribution and kinematics of the massive molecular gas complex 
surrounding a similarly distant unlensed quasar would require a ten fold
improvement both in resolution and sensitivity over the existing
instruments.  Over 20 gravitationally lensed objects suitable for detailed 
studies using this gravitational lens telescope are now known 
in the literature (see \cite{bla92,mao93,kin96}),
and studies exploiting this magnifying characteristic at all wavelengths
may be rewarded with valuable new insights on these distance sources.
 
\acknowledgements

The authors are grateful to E. Falco for kindly providing the HST image
for the analysis and comparison and R. Barvainis for sharing his unpublished
data.  This manuscript also benefitted greatly from the careful reading
and useful suggestions by P. Ho and referee R. Antonucci.
This research is supported in part by NSF Grant AST 93-14079.


\clearpage

\begin{table*}
\centerline{\bf Table 1.  Summary of the Observations}
\begin{center}
\begin{tabular}{lc}
\multicolumn{1}{l}{} &  
\multicolumn{1}{c}{} \\
\tableline
\tableline
RA (B1950)  & $14^h~13^m~20^s.08$ \\
Dec (B1950) & $11^\circ ~43' ~37''.8$ \\
$\nu_{obs}$ &  226.7035 GHz \\
$<z_{CO(7-6)}>$ &  $2.5582\pm0.0003$ \\
Luminosity Distance, $D_L$\tablenotemark{a} & 10.2 $h^{-1}$ Gpc \\
Angular Size Distance, $D_A$\tablenotemark{b} & 0.79 $h^{-1}$ Gpc \\
& ($1''=3.8~h^{-1}$ kpc) \\
$\theta_{FWHM}$  & $0''.86 \times 1''.0$ (PA = 73$^\circ$) \\
$(\Delta V)_{FWHM}$  & $339\pm21$ km s$^{-1}$ \\
$S_{CO} \Delta V$  & $41\pm4$ Jy km s$^{-1}$ \\
$L_{CO(7-6)}$ (=$4\pi D^2_L S\Delta V \nu_{obs}$)  
&  $5.5 \times 10^8~h^{-1}~L_\odot$ \\
$M_{H_2}$  & $2.5 \times 10^{10} h^{-2}~M_\odot$ (for $\mu=10$)\\
$M_{dyn}$  & $2.6 \times 10^{10} [R(kpc)]^{-1} (Sin~i)^{-2} ~M_\odot$\\
\tableline
\tableline
\end{tabular}
\end{center}


\tablenotetext{a}{$D_L = 6000~[(1+z)-(1+z)^{1/2}]~h^{-1}$ Mpc 
(for $\Omega_\circ = 1$)}
\tablenotetext{b}{$D_A = D_L / (1+z)^2$.}



\end{table*}

%
%

\clearpage

\figcaption[apjlfig1.ps]{ 
{\bf (a)} Velocity integrated map of CO (7--6) emission from the
Cloverleaf quasar (H 1413+117) at $0''.86 \times 1''.0$ (PA=73$^\circ$)
resolution in contours overlaid on the gray scale optical image
obtained with the Hubble Space Telescope (Falco 1993).  
The contours are -3, -2, 2, 3, 4, 5, 6, 7, 8, and 9 times the
rms noise (5.3 mJy beam$^{-1}$). 
Image ``A" is brightest in optical light, but the CO peaks near
image ``B", probably reflecting real differences in the
underlying distributions.  The coordinates are in offset with
respect to image A.
{\bf (b)} A difference map between the blue and redshifted 
emission [shown in (c)\&(d)] made in order
to test if the two velocity components are resolved.
The contour levels are the same as in (c)\&(d) for a direct comparison.  
{\bf (c)\&(d)} Blue- and redshifted CO (7--6) emission averaged 
over 145 km s$^{-1}$
are shown in contours superposed on the HST image.  The contours
are -3, -2, 2, 3, 4, 5, and 6 times the rms noise 
(8.3 mJy beam$^{-1}$).  The blue and red images consist of 2-3 
bright peaks, each with distinctly different distribution from the 
other and from the optical light.  While the apparent difference
in morphology between the blue and red image is striking,
only 30\% of the total flux remains in the difference map, suggesting
that the bulk of the flux originates very close to the quasar.
\label{fig1}}

\figcaption[apjlfig2.ps]{
{\bf (a)} Evenly-spaced contours of surface mass 
density in a simple elliptical potential
plus external shear gravitational lens model computed to account for the 
observed locations and relative magnifications of the optical/infrared
images.  The star marks the location of the lensed quasar.  
(The apparent higher brightness of optical image B over A is an artifact of
re-gridding poorly sampled HST image.)
{\bf (b)\&(c)} Blue- and redshifted CO source models produced by
ray-tracing observed CO images through the lens model and 
convolving with a $0''.2$ beam corresponding
to the effective resolution of the gravitational lens.  The 
contours are linear increments of the 15\% of the peak brightness.  
The red and blue CO sources are located roughly on the opposite
side of the quasar, each located $\sim 0''.15$ (550$h^{-1}$ pc) 
away, and the observed distribution may be 
interpreted as the CO emitting region having about 1 kpc in extent with 
a velocity gradient like a rotating disk. \label{fig2}}



\clearpage
\begin{figure}[h]
\vbox to8.5in{\rule{0pt}{8.5in}}
\includegraphics{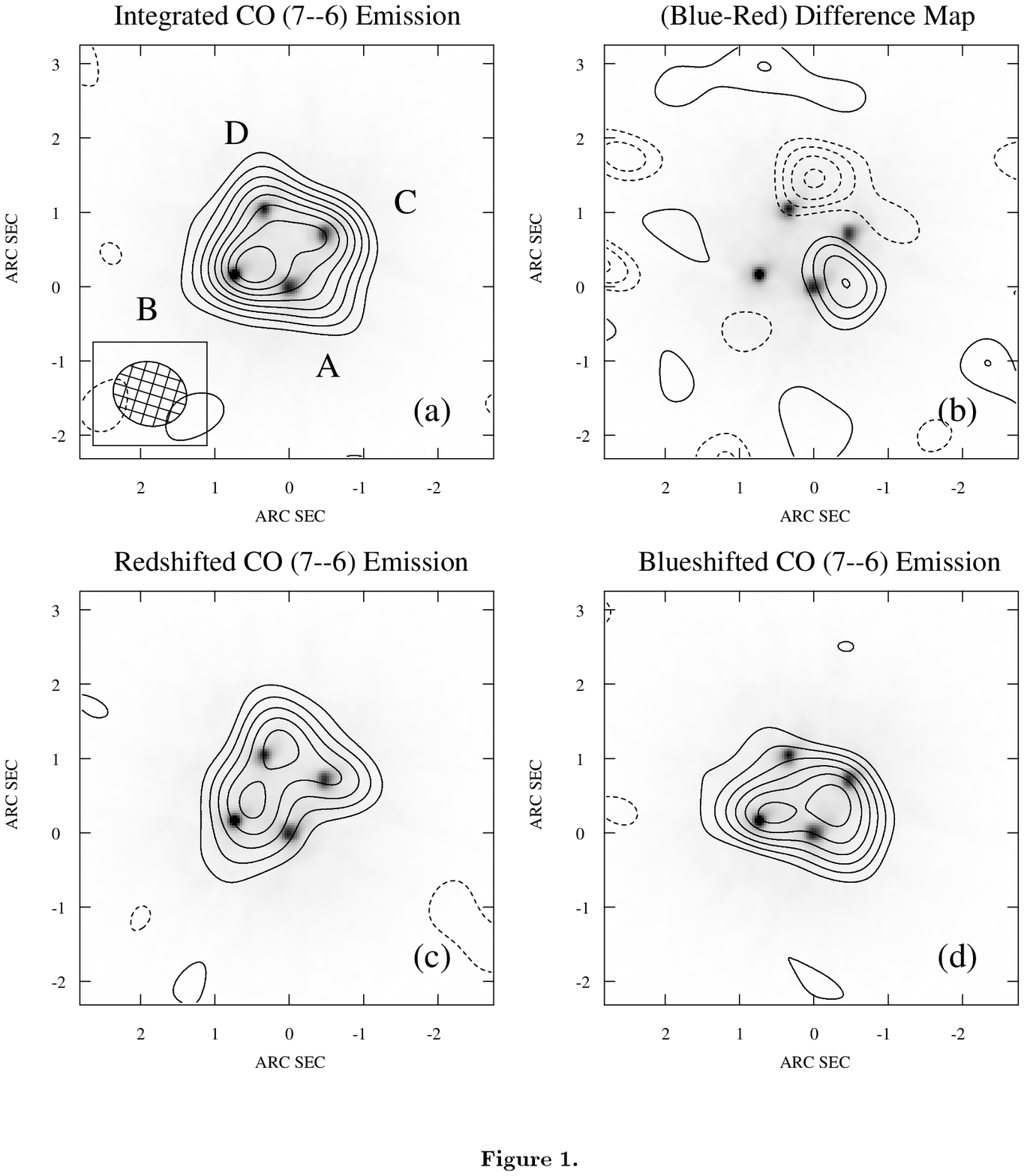}
\end{figure}

\clearpage
\begin{figure}[h]
\vbox to8.5in{\rule{0pt}{8.5in}}
\includegraphics{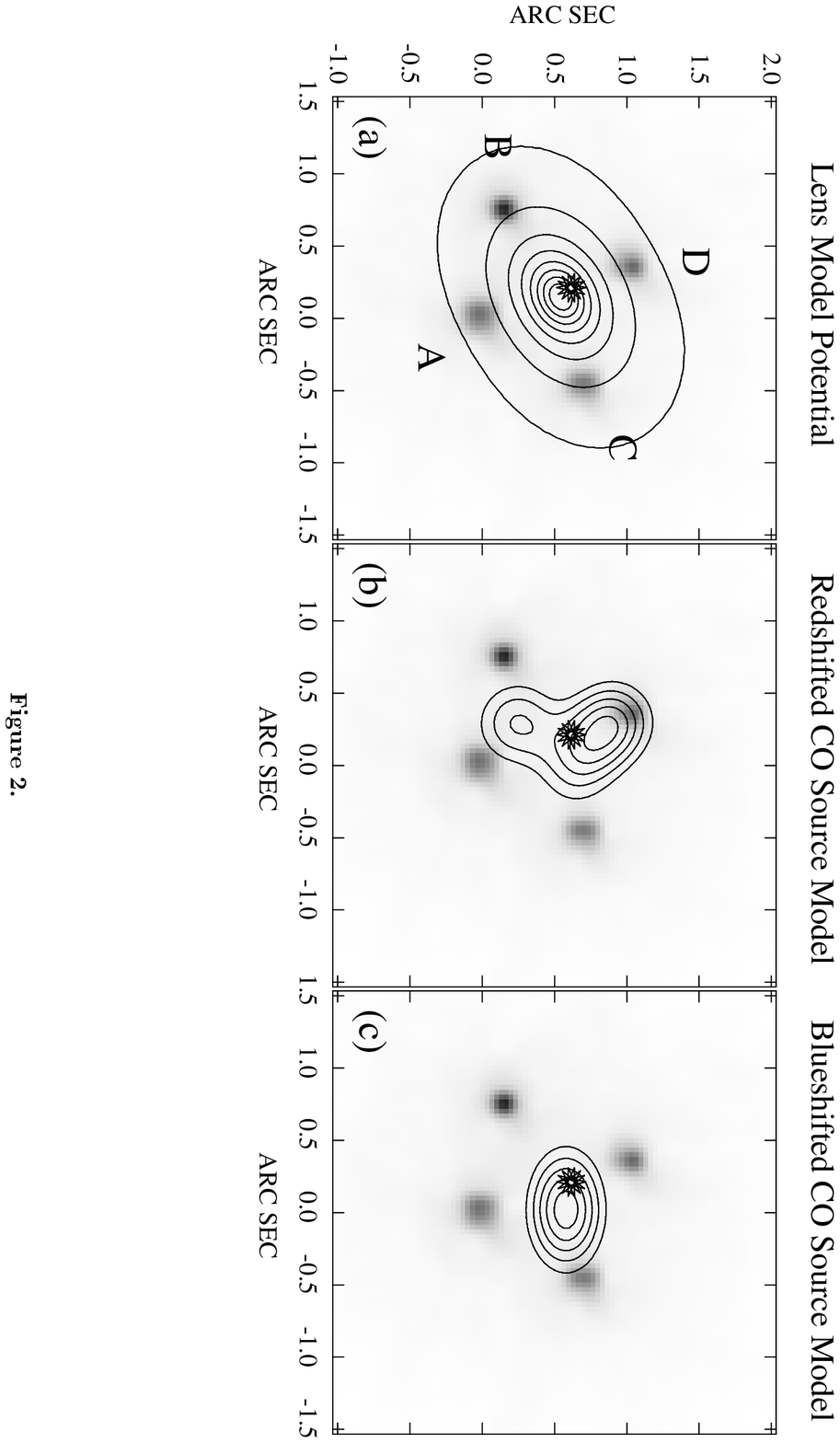}
\end{figure}

\end{document}